%% file: main.tex
\documentclass[sigconf]{acmart}  % ,anonymous
\AtBeginDocument{%
  }

\usepackage[normalem]{ulem}
\usepackage{array}             % for m{width} column specification
\usepackage{balance}

\setcopyright{cc}
\copyrightyear{2025}
\acmYear{2025}
\setcctype{by}
\acmConference[ICMR '25]{Proceedings of the 2025 International Conference on Multimedia Retrieval}{June 30-July 3, 2025}{Chicago, IL, USA}
\acmBooktitle{Proceedings of the 2025 International Conference on Multimedia Retrieval (ICMR '25), June 30-July 3, 2025, Chicago, IL, USA}
\acmDOI{10.1145/3731715.3733422}
\acmISBN{979-8-4007-1877-9/2025/06}

\begin{document}

\title[Robust Relevance Feedback for Interactive Known-Item Video Search]{Robust Relevance Feedback for Interactive \\ Known-Item Video Search}


\author{Zhixin Ma}
\affiliation{
  \institution{Singapore Management University}
  \city{}
  \country{Singapore}}
\email{zxma@smu.edu.sg}

\author{Chong-Wah Ngo}
\affiliation{
  \institution{Singapore Management University}
  \city{}
  \country{Singapore}}
\email{cwngo@smu.edu.sg}

% \renewcommand{\shortauthors}{Ma et al.}
\renewcommand{\shortauthors}{Zhixin Ma and Chong-Wah Ngo}
%% No italics, no superscripts
%% Use footnote or author note to identify equal contribution and/or contact author info

\begin{abstract}
Known-item search (KIS) involves only a single search target, making relevance feedback—typically a powerful technique for efficiently identifying multiple positive examples to infer user intent—inapplicable. PicHunter addresses this issue by asking users to select the top-k most similar examples to the unique search target from a displayed set. Under ideal conditions, when the user's perception aligns closely with the machine's perception of similarity, consistent and precise judgments can elevate the target to the top position within a few iterations. However, in practical scenarios, expecting users to provide consistent judgments is often unrealistic, especially when the underlying embedding features used for similarity measurements lack interpretability. To enhance robustness, we first introduce a pairwise relative judgment feedback that improves the stability of top-k selections by mitigating the impact of misaligned feedback. Then, we decompose user perception into multiple sub-perceptions, each represented as an independent embedding space. This approach assumes that users may not consistently align with a single representation but are more likely to align with one or several among multiple representations. We develop a predictive user model that estimates the combination of sub-perceptions based on each user feedback instance. The predictive user model is then trained to filter out the misaligned sub-perceptions. Experimental evaluations on the large-scale open-domain dataset V3C indicate that the proposed model can optimize over 60\% search targets to the top rank when their initial ranks at the search depth between 10 and 50. Even for targets initially ranked between 1,000 and 5,000, the model achieves a success rate exceeding 40\% in optimizing ranks to the top, demonstrating the enhanced robustness of relevance feedback in KIS despite inconsistent feedback.
\end{abstract}






\maketitle

\input{introduction}
\input{related_work}

\input{problem_statement}

\input{methods}
\input{experiments}

\input{conclusion}

\begin{acks}
This research is supported by the Ministry of Education, Singapore, under its Academic Research Fund Tier 2 (Proposal ID: T2EP20222-0047). Any opinions, findings and conclusions or recommendations expressed in this material are those of the authors and do not reflect the views of the Ministry of Education, Singapore. Professor NGO Chong Wah gratefully acknowledges the support by the Lee Kong Chian Professorship awarded by Singapore Management University.
\end{acks}

% \bibliographystyle{ACM-Reference-Format}
\balance
% \bibliography{reference}
\input{reference.bbl}

\input{appendix}

\end{document}

%% file: introduction.tex
\section{Introduction}
\label{sec:introduction}

Relevance feedback~\cite{rui1997relevancebinary,rui1998relevance5level,zhou2003rfImage,su2003relevance} is widely applied in interactive video retrieval to capture human subjectivity and refine search queries. In cases where there are numerous positive examples within the search space, i.e., the ad-hoc video search, users can contribute by submitting relevant examples or indicating significance ratings to assist the system in aligning with their search goals. Conversely, in known-item search (KIS), where only a unique target item exists, user-labeled positive examples are typically unavailable. PicHunter~\cite{cox1996pichunter, cox1998bayesian, cox2000pichunter}, a Bayesian-based relevance feedback framework for KIS, addresses this by utilizing partially similar, yet not entirely positive, examples to locate the search target. Users provide judgments on which videos are ``closer'' to the target, following a heuristic that videos more similar to selected ones are likely closer to the target than non-selected videos. The probability of relevance is then updated exponentially based on the distance difference between candidate videos and those that have been selected versus non-selected ones. Despite the surge in video volume, the exponential update function effectively refines the search space, improving the target's ranking. Under the assumption that users consistently align with the machine by providing accurate and precise judgments, PicHunter~\cite{cox1996pichunter,cox1998bayesian,cox2000pichunter} is capable of promoting the target to the top position within seven iterations. However, assuming users can consistently make precise judgments is often impractical. To address this, PicHunter implements a ``soft'' approach by iteratively updating a probability distribution over all candidates, thus pushing the search target closer to the top. However, the system still has high sensitivity, as inconsistent user judgments may result in the target being ranked lower in the list.

The rapid growth in video content and the widespread adoption of embedding-based retrieval methods further complicate the application of relevance feedback in KIS. As video datasets expand to cover broader domains, videos conveying similar semantics tend to exhibit unlimited and unpredictable variability in appearance~\cite{smeulders2000cbir}. This variability can lead to inconsistent user judgments, particularly when candidate videos possess subtle, non-comparable differences. Furthermore, the dominance of embedding representations in cross-modality retrieval amplifies challenges in achieving consistent judgments. While explainable features like texture, color, and illumination~\cite{cox2000pichunter} facilitate user judgments, embeddings, designed to align with textual queries, often result in representation distances that are difficult for users to interpret and compare. This mismatch between user perception and machine interpretation can lead to misalignment.

To address these limitations, we propose a novel approach that enhances the robustness and effectiveness of relevance feedback in KIS. Our method decomposes user perception into multiple sub-perceptions, each represented within an independently trained embedding space. Although users may struggle to provide judgments that always align with a single embedding space, they are more likely to agree with one or several of these representations. Accordingly, we develop a predictive user model that estimates the user's perceptual alignment within each embedding space based on the query, the current search state, and the user's interaction history. Specifically, we propose pairwise judgments, allowing users to identify the video closer to the target within a pair, rather than choosing the most similar videos from a larger set. This approach reduces the number of judgments needed per iteration, thereby decreasing the chance of misalignment. 
% Within each feature space, user feedback is represented by the difference between the video pair's features, highlighting content that users deem significant based on the current search state. The user query, a concise video caption generated by multi-modal large language models (LLMs)~\cite{li2023blip2, zhang2024llavanextvideo}, serves as the initial input for a typical relevance feedback scenario. 
To this end, we represent user feedback as the differences between a pair of videos. The differences reflect the feature deviations in their respective sub-perceptions. A user query, generated from a large language model (LLM)~\cite{li2023blip2, zhang2024llavanextvideo}, is a concise caption consisting of up to three sentences narrating the content of a target video.
The search state, represented by the top-ranked videos, approximates the context of the search. Additionally, the predictive model calculates a confidence score for each sub-perception in response to user feedback. This confidence score is incorporated into the Bayesian update, enhancing tolerance for inconsistent judgments compared to binary decisions.

Our display model employs two strategies: greedy sampling and diversity sampling. The greedy strategy follows PicHunter's most-probable sampling approach, drawing pairs from videos with the highest probability of being the target. Since top-ranked videos often share similar semantics or visual characteristics, diversity sampling expands the selection range to prevent overly similar pairs. Training data includes pairs generated by both strategies. During training and evaluation, after the display is visualized, a user simulator provides relative judgments based on majority voting across sub-perceptions. With access to the search target, the user simulator can determine Oracle judgment by comparing their distances to the target within each sub-perception. The ultimate user judgment is simulated by selecting the majority choice among all sub-perceptions. The predictive user model needs to filter out the sub-perception(s) that are misaligned with the simulated judgments.

We evaluate our approach on the large-scale open-domain datasets V3C~\cite{luca2019v3c, berns2019v3c1, rossetto2021v3c2}. To minimize the bias introduced by LLM-generated captions, we select evaluation queries based on the initial rank of the search target, reflecting the difficulty of optimizing the target to the top position through relevance feedback. The initial ranks range from 10 to 5,000, enabling a comprehensive evaluation of the system's capacity to improve search performance across varying levels of difficulty. Experimental results indicate that our approach successfully optimizes over 60\% of queries to rank-1 when initial ranks fall between 10 and 50. For targets initially ranked between 1,000 and 5,000, our method achieves a 40\% chance of reaching rank-1. By incorporating explicit search space pruning tricks, the overall recall@1 performance improves from 0.547 to 0.638, underscoring the enhanced robustness of relevance feedback in KIS. Additionally, we evaluate our method using the textual KIS (t-KIS) query set from the Video Browser Showdown (VBS)~\cite{jakub2021vbs,lokoč2023vbs11,lucia2024vbs12}. Results indicate that our method outperforms the baselines and optimizes up to 16 out of the 17 t-KIS queries to rank-1 with the full t-KIS query text. 
% Although the query encoder faces challenges in processing long queries, the proposed method successfully identifies 58.82\% of queries.

This paper addresses the challenge of aligning user and machine perceptions, a problem often overlooked in large-scale interactive search, to better leverage user feedback for search re-ranking.  Specifically, misinterpreting user perception based on feedback can push a search target farther down the rankings. The main contribution of this paper is the proposal of a predictive model, built upon the classic PicHunter~\cite{cox1996pichunter,cox1998bayesian,cox2000pichunter}, to enhance the robustness of interactive search. By correctly judging user perceptions, search robustness is enhanced by steering the Bayesian update to consistently move the target towards higher rankings.

\vspace{-5px}

%% file: related_work.tex
\section{Related Work}

Relevance feedback has proven particularly effective in the ad-hoc video search by narrowing the search scope, where there are abundant positive examples available. Such methods dynamically train a classifier on-the-fly during the search session to leverage user-labeled relevant items and improve search performance~\cite{yan2003multimedia,kratochvil2020som,lokovc2022video,kratochvil2020somhunter}.
To refine the search space, a fuzzy relevance feedback framework has been proposed, which iteratively optimizes the candidate distribution using positively labeled items from the user~\cite{yap2005soft}. However, these approaches are less suitable for KIS or video moment retrieval tasks~\cite{lei2020tvr,li2020hero,hou2021conquer}, where typically only one target is associated with each query. Although utilizing positive examples to refine the query is impractical in these cases, the relative relationship, for example, which one is more similar to the search target, can still provide useful cues for enhancing search performance.

PicHunter~\cite{cox1996pichunter,cox1998bayesian,cox2000pichunter} proposes a method for iteratively refining the search space by requesting user ``relative judgments'', where the selected items are not strictly positive examples but are instead ``closer'' to the search target than the other items. A Bayesian framework is applied to update the probability of candidate items being positive based on these selections, with the assumption that the chosen items are closer to the target within the representation space. This probability is updated exponentially based on the distances between the selected and non-selected items. However, consistently interpreting images to provide accurate ``relative judgments'' can be challenging, as images are inherently more ambiguous than textual representations. This limitation is further exacerbated in the era of deep learning, where representations are often too complex and less interpretable for users to provide clear relative judgments. Additionally, PicHunter’s display strategy, which is used to select the set of videos for collecting feedback, remains an open issue. While enumerating all possible combinations may yield an optimal solution, such an approach is computationally impractical in real-world applications. SOMHunter~\cite{kratochvil2020somhunter,patrik2021somv2,jakub2023cvhunter} is designed to optimize display selection using a self-organizing map (SOM), which presents the central videos of clusters formed by the SOM algorithm. Same as PicHunter, SOMHunter also struggles to to maintain its effectiveness when user feedback deviates from the system's perception. In addition to the relevance feedback, high-level semantic feedback has also been explored as a means to refine the search space. For instance, in~\cite{yanagi2020interactive}, object-wise question answering is employed to filter and re-rank search candidates, enabling a more informative and proactive approach to candidate refinement. However, the approach lacks scalability and is unable to improve the rankings of search targets that are initially ranked low.

Our method addresses misalignment challenges of relevance feedback within KIS~\cite{cox1996pichunter,cox1998bayesian,cox2000pichunter}. We aim to enhance system robustness by predicting user subjectivity based on their feedback. The subjectivity is decomposed into multiple sub-perceptions and a weighted combination of these sub-perceptions is used to approximate the user's subjective for each feedback.

%% file: problem_statement.tex
\input{figs/pipeline}

\section{Problem Statement}
\label{sec:prob_statement}

As shown in Figure~\ref{fig:pipeline}, a relevance feedback system for KIS typically begins with a text query that provides initialization, followed by interactions among three core components: the display model $f_D(\cdot)$, the user (or user simulator) $f_L(\cdot)$, and the user model $f_P(\cdot)$. To address the misalignment issue, the proposed approach is to predict user perception from the feedback, such that the machine only refines the search space using the sub-perceptions that align with the predicted perception. In other words, the approach strives to align the perceptions between user and machine in every iteration before Bayesian updating and hence can reduce the number of user interactions.

We assume that a machine maintains several sub-perceptions, each corresponding to a distinct feature representation. The problem is to predict which sub-perceptions the user is referencing when providing feedback. As illustrated in Figure~\ref{fig:pipeline}, when the combination of sub-perceptions for the relative judgment is accurately predicted, the Bayesian update in the user model can refine the ranking of the search target $v_T$ aligning with the user’s intent. To achieve this, we frame the user perception estimation as a classification problem. In each iteration, the selection of each sub-perception is treated as a binary classification task for each user judgment, taking into account the initial query $q_0$, the search state, and the interaction history. The search state is represented by the distribution $P=\{p_i\}$, where $p_i$ denotes the probability of the $i$-th candidate being the search target.

%% file: figs/pipeline.tex
\begin{figure*}[!htbp]
    \center
    \begin{minipage}{2. \columnwidth}
        \centerline{\includegraphics[width=0.8 \columnwidth]{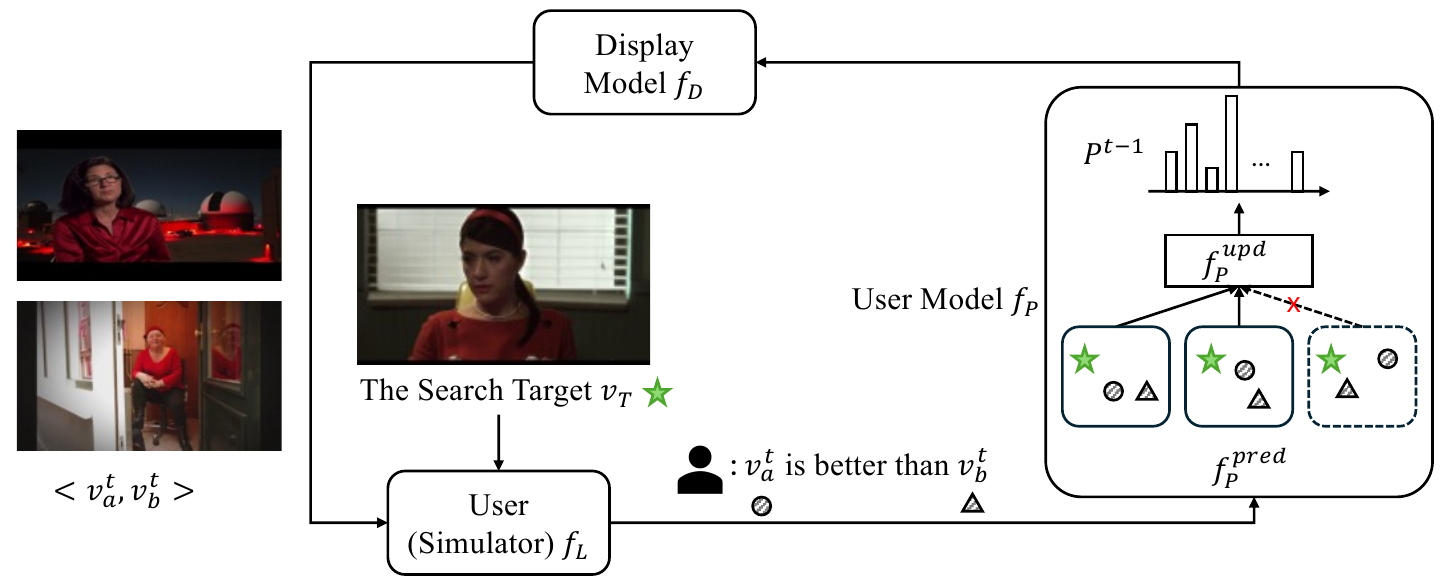}}
    \end{minipage}
    \caption{Framework of the proposed relevance feedback system. The initial query for the search target is ``a woman in a red dress sitting at a desk''.}
    \label{fig:pipeline}
\end{figure*}
\vspace{-5px}

%% file: methods.tex
\section{Methodology}
\label{sec:method}

Pairwise judgment is proposed as the feedback mechanism of the framework shown in Figure~\ref{fig:pipeline}. KIS starts with the user providing a textual query $q_0$ and the machine displays the top-ranked videos for user feedback. In our case, the query $q_0$ is the LLM-generated caption summarizing the search target. The display model visualizes search results for user feedback collection, the user simulator provides relative judgments, and the user model incorporates the user feedback to update the search state from $P^{t-1}$ to $P^t$, where the initial $P_0$ is based on the initial query $q_0$.

\subsection{Pairwise Judgment Feedback}
\label{subsec:pichunter_method_relative}

The interaction with pairwise judgment follows the PicHunter's top-k judgment feedback~\cite{cox2000pichunter}, with some modifications. Specifically, the display $D = \{(v_{a}, v_{b})\}$ is now structured as a list of video pairs, and the user feedback $L = \{l\}$ consists of binary variables. Here, $l$ equals 0 if the user selects $v_{a}$, and 1 if the user selects $v_{b}$. The pairwise judgment feedback provides a more efficient interaction method compared to the top-k feedback in terms of both user cognitive load for providing relative judgments and system computational complexity. 

The top-k feedback mechanism used by PicHunter instructs users to select multiple videos from the recommended videos. Intuitively, the selection can be tedious because, for any video under selection, it involves the similarity comparison to the remaining unselected videos. The selection complexity grows as the number of recommended videos increases. For example, a selection of four videos from a display with size eight can naively involve 22 (i.e., 7+6+5+4) comparisons. When a user selects $m$ videos from $n$ displayed videos, the system needs to calculate the cosine similarity for all combinations of selected ($V_{+}$) and unselected ($V_{-}$) videos, resulting in a complexity of $\mathcal{O}(m \cdot n)$, where $m$ and $n$ are the numbers of selected $V_{+}$ and unselected $V_{-}$ videos, respectively. 
This approach becomes computationally intensive as $n$ and $m$ increase, particularly when updating the probability distribution $P$ on datasets with a large number of videos ($|V|$), leading to potential computational overhead and increased latency in updating the ranking model.

In contrast, pairwise judgment feedback significantly reduces both user effort and computational demands by providing simpler binary comparisons. The user only needs to identify the better video from each pair, resulting in $\frac{n}{2}$ comparisons if there are \(n\) videos in total. From a computational perspective, the system merely needs to calculate the cosine similarity between each pair, resulting in a linear complexity of $\mathcal{O}(n)$. While the reduction in the complexity of computational complexity might not be significant, pairwise judgment has the advantage that it reduces the cognitive load of users by avoiding exhaustive selection of images. The chance that users will make inconsistent judgments is also reduced.

\subsection{Display Model}
\label{subsec:method_display_model}

We adopt pairwise judgment feedback as the interaction method, where users are presented with video pairs and asked to select the video that more closely aligns with the search target in each pair. As discussed in Section~\ref{subsec:pichunter_method_relative}, compared to the PicHunter's Top-k judgment feedback, pairwise judgment feedback reduces the user's cognitive load and the computational complexity in the Bayesian update and has higher robustness to inconsistent user feedback. Figure~\ref{fig:pipeline} illustrates an example of such a video pair.  Given the initial query ``a woman in a red dress sitting at a desk'', the search engine ranks the candidates, pushing the most semantically relevant videos to the top. For example, both the displayed videos depict a sitting woman in a red dress. The display model is then tasked with sampling pairs from the results to collect user feedback.

In our display model, two sampling strategies are introduced: \textit{Greedy Display} and \textit{Diverse Display}. The \textit{Greedy Display} is based on the PicHunter's \textit{Most-Probable} strategy but adapted to accommodate pairwise judgment feedback. Specifically, the greedy display selects $2\times |D|$ videos with the highest probabilities from $P$, where $|D|$ refers to the number of pairs in the display. These videos are then randomly paired to form the display. While selecting the top-ranked videos is reasonable, it often results in pairs of videos with highly similar content, which can make it difficult for users to make distinctions and cause the search to become trapped in a local optimum. 

While PicHunter's another strategy, the \textit{Most-Informative}, can escape local optima by randomly selecting videos from $P$, it does not account for the relationship between the paired videos. To overcome this, we introduce the \textit{Diverse Display}, a modification of PicHunter's \textit{Most-Informative} strategy, which seeks to select the video pairs that are neither too similar nor too dissimilar for user feedback. To prevent the videos in each pair from being irrelevant to the search goal, we consider only the top-$N_{D}$ candidates by probability according to $P$. Then, to increase the diversity between the videos in each pair, one video is sampled from the top 50 candidates (i.e., $[0, 50]$) and the other from the bottom 50 (i.e., $[N_{D}-50, N_{D}]$). Both display strategies are employed during the training to augment the training data, while during the evaluation, only the greedy strategy is used to sample the display.

\subsection{User Simulator}
\label{subsec:user}

When presented with the display, the user directs the search by providing relative judgments. During both training and testing of the model, a user simulator is needed to simulate human behavior. In PicHunter, a perfect user is assumed, capable of making judgments that align perfectly with the system's similarity measure based on video representations. More specifically, both the user and system are assumed to have a single fully aligned perception throughout the period when multiple rounds of feedback are provided for a query. However, this assumption is unrealistic, particularly when the video features are represented by embeddings, which are less interpretable to human users compared to more explainable features like color. Our approach seeks to improve the system's robustness to a situation in which the user provides multiple rounds of feedback, each possibly reflecting a different perception. We call this situation as the imperfect user scenario. In our method, we acknowledge that the perfect user scenario is impractical but assume that a user perception is likely to align with one or a combination of several representations (or sub-perceptions) defined by the system.

We decompose user perception into multiple sub-perceptions, each represented by an embedding feature, denoted as $f_i$. For video pairs in the display, the simulated user makes the relative judgments independently for each sub-perception by selecting the video that is closer to the search target based on the video representation corresponding to that perception. The final decision of the simulated user is made through the majority voting across all sub-perceptions. For instance, for a video pair $(v_a, v_b)$, the user simulator would select $v_a$ in their feedback if more than half of the sub-perceptions determine that $v_a$ has a higher similarity to the search target $v_T$ than $v_b$. Consequently, the user feedback at time $t$ is structured as a list of binary variables $L^t=\{l^t\}$, where the value $0$ of $l^t$ denotes the selection of $v_a$; otherwise, $v_b$ is labeled as the video closer to the target.

Since these sub-perceptions are also used to update the probability $P$ in the user model, inconsistency arises if not all sub-perceptions choose the same video. In such cases, the sub-perceptions that do not agree with the results of the majority decision are treated as misalignment, reflecting the imperfect user scenario. The system's search performance degrades if these misaligned sub-perceptions are used to update $P$. To address this, a user model that includes a module to predict and filter out misaligned sub-perceptions is proposed in the following section.

\input{figs/user_model_predict}

\subsection{Predictive User Model}
In this section, we propose the predictive user model that comprises two modules: the user perception prediction module $f^{\text{pred}}_P$ and the update module $f^{\text{upd}}_P$. In PicHunter, where a perfect user is assumed, the user model only includes an update module that performs a Bayesian update based on all user judgments without any filtering. In contrast, our approach incorporates the prediction model $f^{\text{pred}}_P$ to handle the imperfect user setting by predicting the combination of sub-perceptions for each user feedback. Rather than a binary selection, each sub-perception is assigned a confidence score $c_i$. This confidence score is then used in the Bayesian update of $f^{\text{upd}}_P$ to weight the contribution of each sub-perception during the judgment process, enabling a soft update, which increases the system's tolerance to user feedback of varying perceptions.

\subsubsection{User Perception Prediction}
\label{subsubsec:method_user_perception_prediction}
A prediction model $f_P^{pred}$ is trained independently for each of the sub-perceptions. At each iteration $t$, the model $f_P^{\text{pred}}$ estimates the confidence score $c_i$ based on the user feedback ($D^t, L^t$), historical interactions ($D^1, L^1, \dots, D^{t-1}, L^{t-1}$), and the search state. A search state comprises the videos and their probability distribution $P^t=\{p^t_i\}$. Empirically, we only consider the top 50 videos with the highest probability $p_i$. We represent the search state as an embedding, which is obtained by mean pooling the representations of top-50 videos. 

For the user feedback ($D^t, L^t$), the feature difference between the selected and non-selected videos in a pair is used to represent the relative judgment. Denote $v^+$ and $v^-$ as the representations of the selected and unselected videos, respectively. A relative judgment is calculated as $v_{\text{diff}} = v^+ - v^-$. To clarify the role of $v_{\text{diff}}$, we can compare it to a term in the formulation of Bayesian update. In PicHunter, the probability of the $i$-th video $p_i^t$ at time $t$ is updated iteratively as: $p_i^t = p_i^{t-1} \cdot \hat{p}_i^t$, where $\hat{p}_i^t$ represents the temporal probability calculated based on the relative judgment received at iteration $t$.
The calculation of the temporal probability has been updated as follows: 
\begin{equation}
  \label{eq:pairwise_bayesian}
\hat{p}^t_i = \sum_{(v^{+}, v^{-})} \frac{1}{1 + EXP(-\frac{1}{\rho} \left[s(v^{+}, v_i) - s(v^{-}, v_i)\right])}
\end{equation}
where $s(\cdot)$ denote the cosine similarity and $v_i$ denotes the $i$-th video in the search space. 
Now we discuss the relationship between the feature difference $v_{\text{diff}}$ and the video $v_i$. The similarity between $v_{\text{diff}}$ and $v_i$ is calculated as: 
\begin{align}
     s(v_{\text{diff}}, v_i) & =  \frac{v_{\text{diff}} \cdot v_i}{||v_{\text{diff}}||} \\
     & = \frac{(v^+ - v^-) \cdot v_i}{||v_{\text{diff}}||} \\
     & = \frac{1}{||v_{\text{diff}}||} \cdot \left[s(v^+, v_i) - s(v^-, v_i)\right]
\label{eq:diff}
\end{align}
As shown in Equation~\ref{eq:diff}, the similarity between $v_{\text{diff}}$ and $v_i$ 
is proportional to the term $s(v^+, v_i) - s(v^-, v_i)$ in the Bayesian update.

% By moving around the terms in Equation~\ref{eq:diff}, we have $s(v^+, v_i) - s(v^-, v_i) = ||v_{\text{diff}}|| \cdot s(v_{\text{diff}}, v_i)$. Then, the denominator in equation~\ref{eq:pairwise_bayesian} can be rewritten as $1 + \left[EXP(-\frac{1}{\rho}\, s(v_{\text{diff}}, v_i))\right]^{||v_{\text{diff}}||}$. The denominator's value is sensitive to the value of $||v_{\text{diff}}||$. The sensitivity increases when $||v_{\text{diff}}||$ is close to $0$, which usually happens in the later rounds of interaction, where the top-ranked videos get more and more similar. Although the predictive model can capture the $s(v_{\text{diff}}, v_i)$ from inputs, it can hardly to sense the subtle change in $||v_{\text{diff}}||$.

By rearranging the terms in Equation~\ref{eq:diff}, we obtain $s(v^+, v_i) - s(v^-, v_i) = \|v_{\text{diff}}\| \cdot s(v_{\text{diff}}, v_i).$ Substituting this into the denominator of Equation~\ref{eq:pairwise_bayesian}, we rewrite it as $1 + \left[\exp\left(-\frac{1}{\rho} \, s(v_{\text{diff}}, v_i)\right)\right]^{\|v_{\text{diff}}\|}.$ The value of the denominator is sensitive to changes in $\|v_{\text{diff}}\|$, particularly when $\|v_{\text{diff}}\|$ approaches zero. This typically occurs in later rounds of interaction, where top-ranked videos become increasingly similar. While the predictive model can effectively capture $s(v_{\text{diff}}, v_i)$ from the inputs, it struggles to detect subtle variations in $\|v_{\text{diff}}\|$. To address this, we introduce the distance embedding $E^d$. Specifically, we divide the possible range of $||v_{\text{diff}}||$, i.e., (0, 1], into 100 intervals and assign a learnable embedding to each interval, allowing the model to capture subtle variations in the L2-norm of the feature differences. The impact of distance embedding is analyzed in the ablation study.

As depicted in Figure~\ref{fig:architecture}, a transformer-based model takes the initial query $q_0$, along with both the current and historical user judgments as input, followed by a fully connected layer to output the confidence score $\{c_i\}$ for each sub-perception. Noted that user judgment is modeled as the sum of $v_{\text{diff}}$, $v_{s^t}$ and $E^d$, where $v_{s^t}$ is the embedding of search state at iteration $t$. In summary, the confidence scores $c_i$ are computed as follows:
\begin{equation*}
    \{c_i^t\}_{i=1}^{|D|} = f_P^{\text{pred}}([q_0;D^1,L^1,v_{s^1},\dots,D^{t-1},L^{t-1},v_{s^{t-1}};D^t,L^t,v_{s^t};E^d])
\end{equation*}
The model parameters are optimized using binary cross-entropy loss, which will be elaborated in Section~\ref{subsec:model_learning}.

\subsubsection{Soft Bayesian Update}

The update model $f_P^{\text{upd}}$ incorporates the model-aligned user feedback to progressively refine the search state by updating the probability distribution from $P^{t-1}$ to $P^{t}$. The initial probability distribution $P^0$ is derived from the initial query $q_0$. At each iteration, the probability $p_i$ for the $i$-th video is updated as follows:
\begin{equation}
    \label{eq:bayesian_all}
    p_i^t = p_i^{t-1} \cdot \hat{p}_i^t = \prod_{k=0}^t \hat{p}_i^k
\end{equation}
where $\hat{p}_i^{t}$ represents the temporal probability of the $i$-th video candidate at interaction $t$. The $\hat{p}^t_i$ aggregates the model-aligned user judgments from all the sub-perceptions together, calculated as $\hat{p}^t_i = \sum_{f} \hat{p}^{t,f}_i$, where $f$ denotes the index of the sub-perception. In the $f$-th sub-perception, $\hat{p}_i^{t,f}$ incorporates the confidence score $c$ to weight the contribution of the pairs, obtained as follows:
\begin{equation}
    \label{eq:bayesian_soft}
    \hat{p}^{t,f}_i = \sum_{(v^{+}_j, v^{-}_j)} \frac{1}{1 + EXP(-\frac{1}{\rho} \cdot c_j \cdot \left[s^f(v^{+}_j, v_i) - s^f(v^{-}_j, v_i)\right])}
\end{equation}
where $s^f(\cdot)$ denotes the cosine similarity calculated based on the feature representation of the $f$-th sub-perception.

\subsubsection{Search Space Pruning}
Although the initial query cannot guarantee to locate the search target, it helps to initialize a preliminary ranked list where the content of top-ranked video shots is close to the search target. However, the subsequent user feedback might boost the ranks of irrelevant videos, which are ranked low in the initial search list. As shown in Figure~\ref{fig:method_prune}, the query ``a man is walking down the street with a backpack'' is employed as the initial query. From the top-ranked results, a pair of shots is selected and displayed for user feedback. Compared to the man in red who wears a single-shoulder bag in the top shot, the one in the bottom shot has a similar backpack as the target and wears a dark coat. With the user feedback, the rank of the search target is improved from 244 to 62. However, irrelevant shots showing a backpack and dark coat are also elevated, for example ``a woman walks on a street with a backpack'', ``a man carrying backpack rides a bike'', ``a man in black walks through a corridor with a backpack'', and ``a woman stands in the snow with a backpack''. 
% The elevation of the irrelevant videos is likely to make the displayed pairs incomparable: users cannot select the ``closer'' one from two irrelevant videos.

We engineer the problem by considering only the top-ranked video shots retrieved for the initial query for display. Specifically, we define a threshold, $N_{prune}$, representing the number of video shots for Bayesian update. Candidates outside the top-$N_{prune}$ are excluded from being selected for display.  Due to the reduction in search space, the videos updated with higher probability by Bayesian update are also less likely to drift from the initial query. The embedding of the current search state is also stable, not capturing diverse content irrelevant to the initial query.

\input{figs/method_prune}

\subsection{Model Learning}
\label{subsec:model_learning}
We use binary cross entropy (BCE) as the loss function to characterize the mismatch between the predicted and actual user perceptions. The BCE loss function is defined as follows:
\begin{equation}
L_{BCE} = -\frac{1}{N} \sum_{i=1}^{N} \left[ y_i \log(c_i) + (1 - y_i) \log(1 - c_i) \right],
\end{equation}
where $N$ is the number of feedbacks, $y_i$ represents whether a feedback aligns with the sub-perception, and $c_i$ is the predicted confidence score. The BCE losses of different sub-perceptions are then averaged. 

We perform self-supervised learning by using the simulated interactions in the perfect user setting. Denote an interaction trajectory as $[q_0;D^1,L^1,v_{s^1},\dots,D^{t-1}, L^{t-1},v_{s^{t-1}};D^t,L^t,v_{s^t}]$, where the search target is reached at $t$th iteration. We simulate the system to generate a large number of interaction trajectories. Empirically, we store the trajectories whose $t<=7$ as the training examples, such that the model is trained using these examples to minimize the number of user interactions. Using the trajectories, the model is trained to predict the sub-perception of every user feedback at iteration $t$ in the imperfect user setting.

%% file: figs/user_model_predict.tex
\begin{figure*}[!htbp]
    \center
    \begin{minipage}{2. \columnwidth}
        \centerline{\includegraphics[width=.9 \columnwidth]{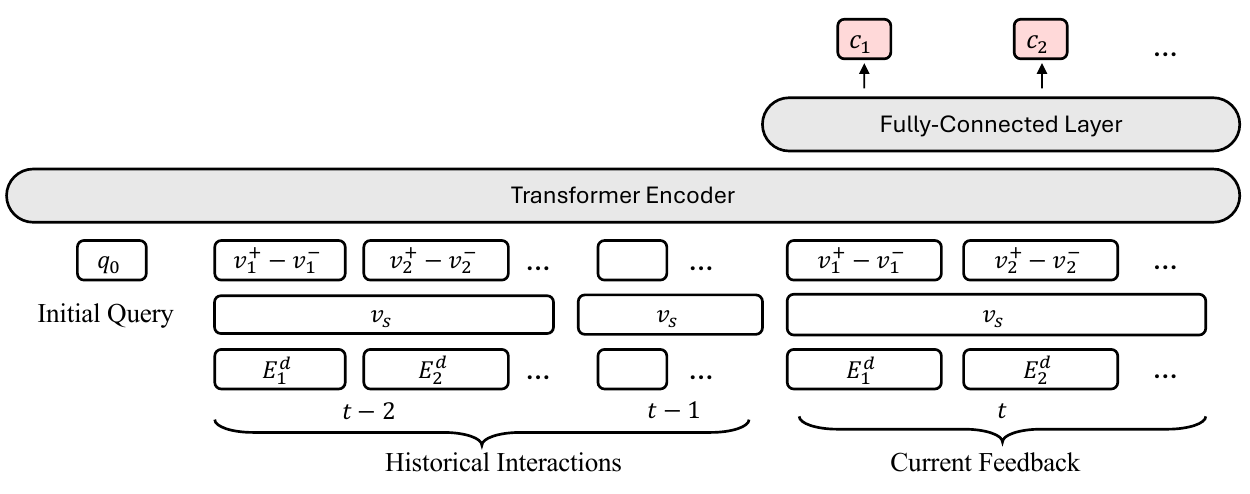}}
    \end{minipage}
    \caption{Architecture of the proposed predictive model $f_P^{pred}$. For simplicity, we annotate the time step at the bottom instead of on the symbols.}
    \label{fig:architecture}
\end{figure*}
\vspace{-10px}

%% file: figs/method_prune.tex
\begin{figure}[!htbp]
    \center
    \begin{minipage}{1. \columnwidth}
        \centerline{\includegraphics[width=1. \columnwidth]{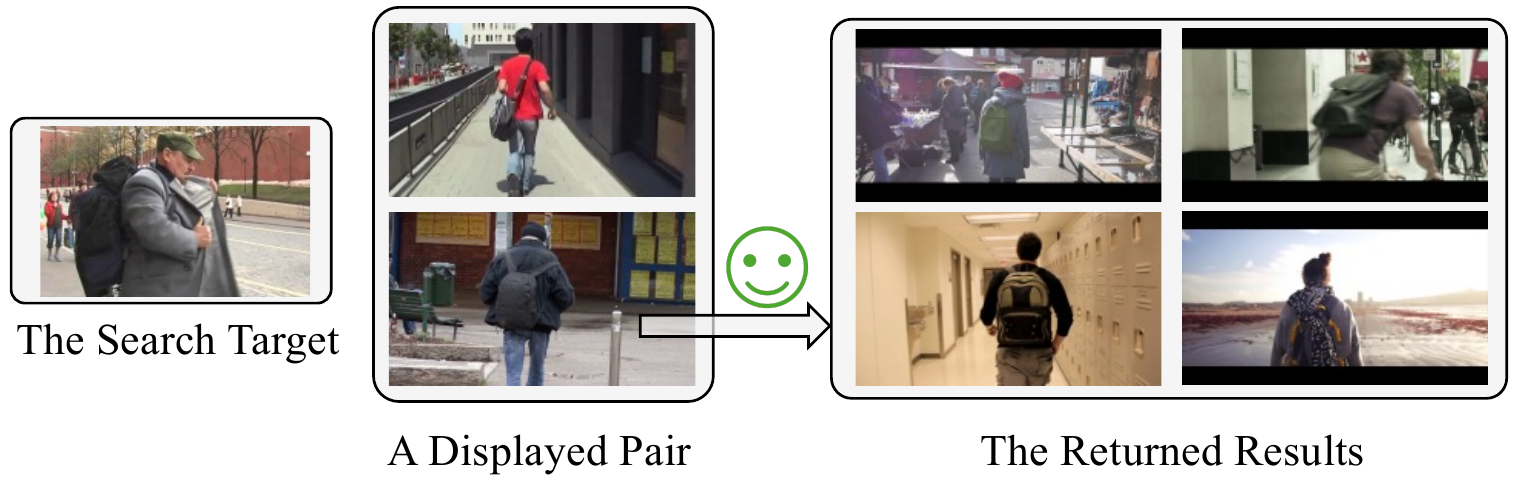}}
    \end{minipage}
    \caption{Illustration of the shift in search results. The initial query is ``A man is walking down the street with a backpack''. The search target is boosted to 62th rank from 244th rank after the second shot in the display panel is selected as feedback. Nevertheless, the ranks of irrelevant images, which are similar to the selected shot but not the query, are also elevated.}
    \label{fig:method_prune}
\end{figure}
\vspace{-10px}

%% file: experiments.tex
\section{Experiments}
We begin this section by introducing the dataset and experiment settings. As different ways of expressing a query can impact retrieval performance, we also compare the effect of using different LLMs for generating captions as queries. Finally, we present the performance of the proposed approach in comparison to two variants of PicHunter.

\subsection{Dataset and Setting}
The experiments are conducted on the large-scale open-domain datasets, V3C1~\cite{berns2019v3c1} and V3C2~\cite{rossetto2021v3c2}, which consist of 1,082,659 and 1,425,454 video clips, respectively. The videos in V3C1 are used as queries to train the user model for the prediction of sub-perceptions, while V3C2 is used for performance evaluation. Note that the testing set is formed by all the videos in V3C2, and the testing queries are sampled from V3C2 as elaborated in Section~\ref{subsec:exp_llm_captions}. We use CLIP4Clip~\cite{luo2021clip4clip}, ITV~\cite{wu2020interpretable}, and BLIP~\cite{li2022blip} features as three sub-perceptions. We limit the number of user interactions to up to seven iterations. In the display model, the number of pairs in display $D$ is set to 5. For the strategy of \textit{diverse display}, $N_D$ is set to 100, where the pair is formed by a video sampled from the top-50 and a video from the bottom-50 of the top-$N_D$ candidates. In the user model, the hyper-parameter $\rho$ used in the Bayesian update is set to $0.05$ and the threshold for search space pruning $N_{prune}^t$ is set to 5,000.  The textual queries are the captions generated from the query videos. We employ BLIP2~\cite{li2023blip2} and LLaVA-NeXT-Video~\cite{zhang2024llavanextvideo} for video captioning.
The effect of the caption sources will be elaborated in Section~\ref{subsec:exp_llm_captions}.

\subsection{Multi-modal LLMs Generated Captions}
\label{subsec:exp_llm_captions}

We use LLaVA-NeXT-Video~\cite{zhang2024llavanextvideo} and BLIP2~\cite{li2023blip2} to generate captions for the target shots, serving as the user query to assign the initial probability to the search candidates. Although the LLaVA-NeXT can generate detailed and precise video captions, we limit the LLM to generating a concise caption to simulate the human query. We use ``Please provide a video description with two or three sentences and keep it as brief as possible, focusing on the main subjects, their actions, and the background scenes.'' as the captioning prompt. Figure~\ref{fig:caption_example} shows examples of captions generated by LLaVA-NeXT and BLIP2. To assess the bias introduced by the captioning model, we sample an evaluation set composed of queries where the search targets reside at different depths of their search rank lists. The challenge of boosting the search targets to the top-ranked position is basically proportional to the search depth. Specifically, we select 4,000 search targets from each of the following rank intervals: (10, 50], (50, 100], (100, 500], (500, 1,000], and (1,000, 5,000], resulting in a total of 20,000 targets for evaluation. Note that there are two sets of 4,000 search targets, one from using LLaVa-NeXTcaptions as queries and the other from BLIP2.

\input{figs/caption_example}

Figure~\ref{fig:exp_caption_res} demonstrates the Recall@1 performance of video caption queries from two different LLMs, LLaVA-NeXT-Video~\cite{zhang2024llavanextvideo} and BLIP2~\cite{li2023blip2}, across different ranges of the target's initial rank. The x-axis denotes the range of depth where a search target resides in its rank list. The y-axis shows the recall@1 performance up to seven rounds of user interactions. The results are averaged over all the queries that fall in a particular search range.

The results show a marginal difference between BLIP2 and LLaVA-NeXT-Video across different ranges of search depth. For example, in the 10-50 range, BLIP2 slightly surpasses LLaVA-NeXT-Video, but in the subsequent ranges, such as 50-100 and 100-500, both models exhibit nearly identical performance. This suggests that the choice of the captioning model (whether LLaVA-NeXT-Video or BLIP2) has minimal impact on the retrieval success after several iterations of the search.

\input{figs/caption_comparison}

\input{tabs/exp_main}

\subsection{Performance Comparison}

Table~\ref{tab:exp_main} compares the Recall@1 performance of three models-\textit{Random}, \textit{PicHunter}, and our proposed model (denoted as ``\textit{Ours}'') over seven iterative steps. The user model \textit{Random}, different from the random strategy introduced in the display model, randomly selects sub-perceptions for each user feedback to update the probability distribution $P$. In the \textit{PicHunter} baseline, the user model assumes that relative judgments are aligned with all sub-perceptions, updating $P$ without filtering out inconsistent sub-perceptions. In contrast, \textit{Ours} updates $P$ and optimizes the target's rank with the model-aligned perception. The table highlights the robustness of each system, as reflected in their ability to recall the correct result at rank 1. The Random model demonstrates the lowest performance, with only a slight improvement across the steps, starting from around 0.02 at Step 1 and gradually increasing to 0.23 at Step 7. The PicHunter baseline, while better than Random, shows stronger improvements, beginning at 0.0187 and achieving 0.49 by Step 7.

The proposed model consistently outperforms both baselines at each step. It starts at 0.0238, which is higher than both Random and PicHunter at Step 1, and continues to surpass them in every step. By Step 7, the proposed model achieves a Recall@1 of 0.5467, representing the highest robustness among the three models. The progressive gap between the proposed model and the baselines widens as the iterations increase, suggesting that the proposed system is more efficient and effective at learning from previous iterations and improving retrieval accuracy. The significant improvement of the proposed model over the baselines, particularly PicHunter, underscores its superior performance and robustness in video retrieval tasks.

Figure~\ref{fig:exp_rangewise} presents the Recall@1 performance comparison among these three approaches for the search targets residing at different depths. The Random baseline consistently underperforms two other approaches across all ranges of search depth. For example, in the 10-50 range, it achieves a recall of approximately 0.3, and the performance gradually drops thereafter. In the 1000-5000 range, the Random baseline recall drops below 0.1, reflecting its struggle with more difficult retrieval tasks. PicHunter significantly improves over Random, particularly in the lower rank ranges. For instance, in the 10-50 range, PicHunter achieves a recall of approximately 0.58, and in the 50-100 range, its performance achieves around 0.53. However, as the rank range becomes more challenging (e.g., 500-1000 and 1000-5000), PicHunter's performance also declines, though it consistently outperforms the Random baseline. The proposed model, however, consistently outperforms both baselines across all difficulty levels. In the 10-50 range, it reaches nearly 0.65 recall, which is considerably better than PicHunter. Similar to PicHunter, the performances also drop proportionally with the increase in search depth. However, recall@1 is consistently better than PicHunter across all search depths.

\input{figs/rangewise_results}

We conduct a significance test by a paired t-test on each search depth to contrast the performances of Ours and PicHunter. At the significance level of 0.05, with a p-value of 0.0114, the results verify that Ours is significantly better than PicHunter across all depths.

\subsection{Search Space Pruning}

In this section, we perform explicit pruning of the search space. Explicit pruning filters out irrelevant videos, thereby preventing the probabilities of irrelevant candidates from increasing due to ambiguity in relevance feedback. Specifically, we discard candidates that are initially ranked lower than 5,000 based on the textual query. In Table~\ref{tab:exp_prune}, we compare Recall@1 performance at different steps for various models, with and without search space pruning. The baseline model, PicHunter, and the proposed model, labeled as ``Ours'' are evaluated across seven iterative steps. The proposed model consistently outperforms PicHunter at each step. Furthermore, the table also includes variants of these models that employ search space pruning, indicated as ``PicHunter w/ Prune'' and ``Ours w/ Prune''. The use of search space pruning leads to notable performance gains for both models. For instance, after 7 steps, PicHunter achieves a Recall@1 value of 0.4949, while ``PicHunter w/ Prune'' reaches 0.6234. Similarly, the proposed model without pruning attains 0.5467, whereas the variant with pruning, ``Ours w/ Prune'' achieves the highest score of 0.6384. Across all steps, the pruning technique enhances recall performance, with ``Ours w/ Prune'' demonstrating the best results, particularly evident in the highest Recall@1 value among all models after 7 steps. This suggests that search space pruning is consistent and highly effective in improving retrieval performance.

\input{tabs/exp_prune}

\subsection{Performance on VBS textual-KIS Query}
\label{sec:exp_tkis}

% (See Table~\ref{tab:tkis_text} in the supplementary document)

We evaluate the proposed model and baselines on the V3C2 dataset using the 17 textual KIS (t-KIS) queries~\footnote{Please refer to the query text in the supplementary document.} 
from video browser showdown (VBS)~\cite{jakub2021vbs,lokoč2023vbs11,lucia2024vbs12} 2022 to 2024. In the t-KIS task, a query is presented in three rounds, with more details being supplemented to the query in each round. Table~\ref{tab:exp_tkis} shows the number of queries where their search targets are successfully identified over three different rounds. Note that the search target may involve multiple shots as defined in the V3C2 dataset. In such cases, we evaluate recall performance by considering the best rank among all involved video shots.

As shown in Table~\ref{tab:exp_tkis}, the proposed model resolves the most number of queries in round 1, achieving the highest recall@1 and recall@10. With more details being exposed for these queries in the next two rounds, our model manages to rank the search targets for 16 out of 17 queries at top-1 and all queries within top-10. Compared to Random and PicHunter, our model not only finds more search targets when reaching the 3rd round but also the fastest in terms of number of iterations required to find these queries. For example, our model requires 4.5 iterations to rank a target to the top-1 position on average, versus Random and PicHunter which require 5.25 and 4.57 iterations, respectively. In terms of speed, our model needs 0.067 seconds for each iteration, which is slower than 0.025 seconds by PicHunter. Nevertheless, the slower speed is compromised by search efficiency and fewer iterations. Overall, the interaction speed is in real-time for all three compared approaches.

\input{tabs/exp_tkis_query}

%% file: figs/caption_example.tex
\begin{figure}[!htbp]
    \center
    \begin{minipage}{1. \columnwidth}
        \centerline{\includegraphics[width=.9 \columnwidth]{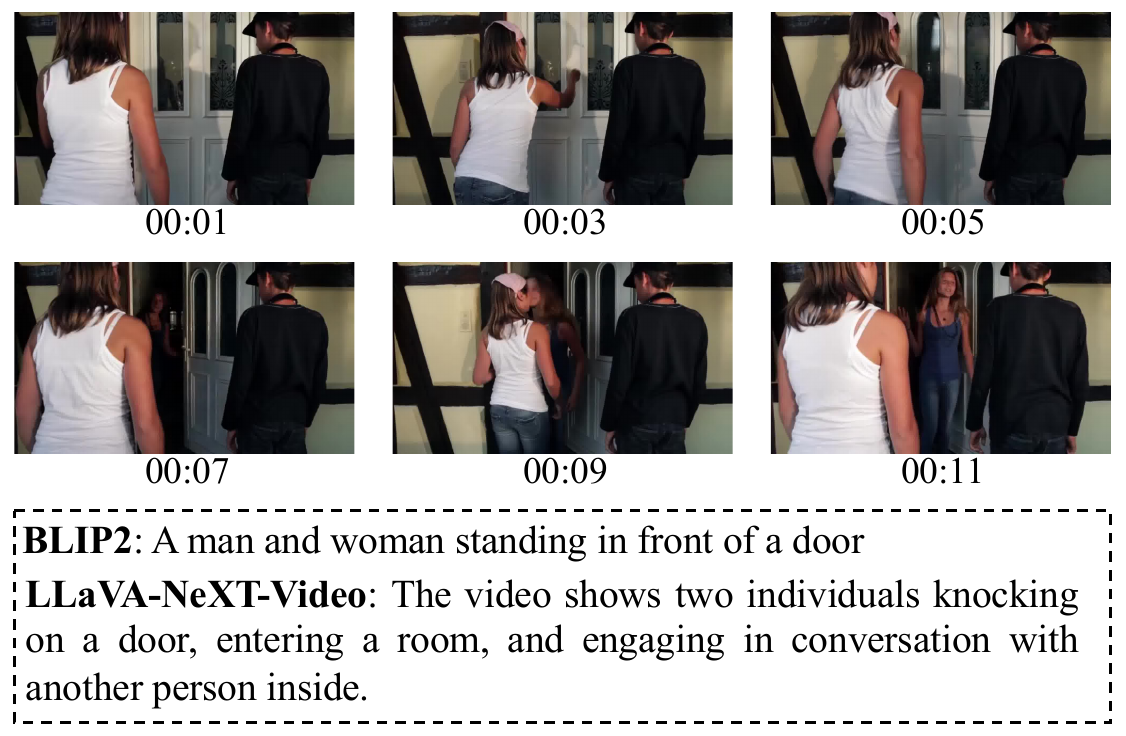}}
    \end{minipage}
    \caption{Example of LLaVA-NeXT and BLIP2 captions.}
    \label{fig:caption_example}
\end{figure}

%% file: figs/caption_comparison.tex
\begin{figure}[!htbp]
    \center
    \begin{minipage}{1. \columnwidth}
        \centerline{\includegraphics[width=1. \columnwidth]{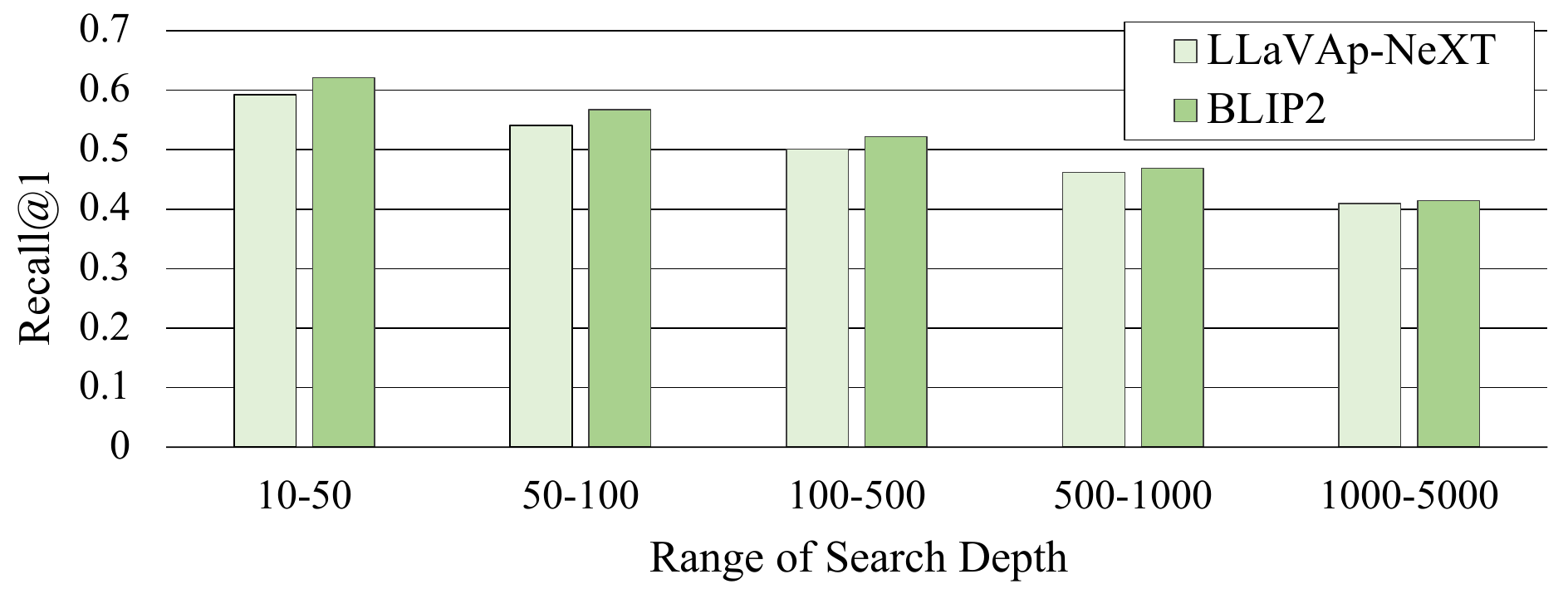}}
    \end{minipage}
    \caption{Performance comparison of different caption sources.}
    \label{fig:exp_caption_res}
\end{figure}
\vspace{-10px}

%% file: tabs/exp_main.tex
%%% CIB
\begin{table*}[h]
\caption{Recall@1 performance comparison with baselines at different steps.}
\label{tab:exp_main}
\centering
\begin{tabular}{lccccccc}
\toprule
            & Step-1    & Step-2          & Step-3          & Step-4          & Step-5          & Step-6          & Step-7          \\ \hline
Random      & 0.0155    & 0.0355          & 0.0624          & 0.0946          & 0.1351          & 0.1795          & 0.2308          \\
PicHunter & 0.0187    & 0.0507          & 0.1091          & 0.1874          & 0.2861          & 0.3942          & 0.4949          \\ 
Ours        & \textbf{0.0238}    & \textbf{0.0688} & \textbf{0.1409} & \textbf{0.2363} & \textbf{0.3438} & \textbf{0.4496} & \textbf{0.5467} \\ 
\hline \bottomrule
\end{tabular}
\end{table*}

\vspace{-10px}

% %%% CCsEv
% \begin{table*}[h]
% \caption{Recall@1 performance comparison with baselines at different steps.}
% \label{tab:exp_main}
% \centering
% \begin{tabular}{lccccccc}
% \toprule
%             & Step-1   & Step-2   & Step-3   & Step-4   & Step-5   & Step-6   & Step-7   \\ \hline
% Random      & 0.0127   & 0.023    & 0.0351   & 0.0493   & 0.0664   & 0.0868   & 0.1084   \\
% PicHunter   & 0.0141   & 0.0347   & 0.0617   & 0.0982   & 0.1384   & 0.1817   & 0.2268   \\
% Ours        & \textbf{0.0253}   & \textbf{0.065}   & \textbf{0.1219}   & \textbf{0.1926}   & \textbf{0.2698}   & \textbf{0.3462}   & \textbf{0.4134}   \\ \hline
% \bottomrule
% \end{tabular}
% \end{table*}

% %%% CCs
% \begin{table*}[h]
% \caption{Recall@1 performance comparison with baselines at different steps.}
% \label{tab:exp_main}
% \centering
% \begin{tabular}{lccccccc}
% \toprule
%             & Step-1   & Step-2   & Step-3   & Step-4   & Step-5   & Step-6   & Step-7   \\ \hline
% Random      & 0.0114   & 0.0213   & 0.0306   & 0.044    & 0.0604   & 0.0814   & 0.1041   \\
% PicHunter   & 0.0163   & 0.0381   & 0.0685   & 0.1104   & 0.157    & 0.2056   & 0.2497   \\
% Ours        & \textbf{0.0278}   & \textbf{0.0713}   & \textbf{0.1386}   & \textbf{0.2286}   & \textbf{0.3332}   & \textbf{0.4341}   & \textbf{0.528}   \\ \hline
% \bottomrule
% \end{tabular}
% \end{table*}

%% file: figs/rangewise_results.tex
\begin{figure}[!htbp]
    \center
    \begin{minipage}{1. \columnwidth}
        \centerline{\includegraphics[width=1. \columnwidth]{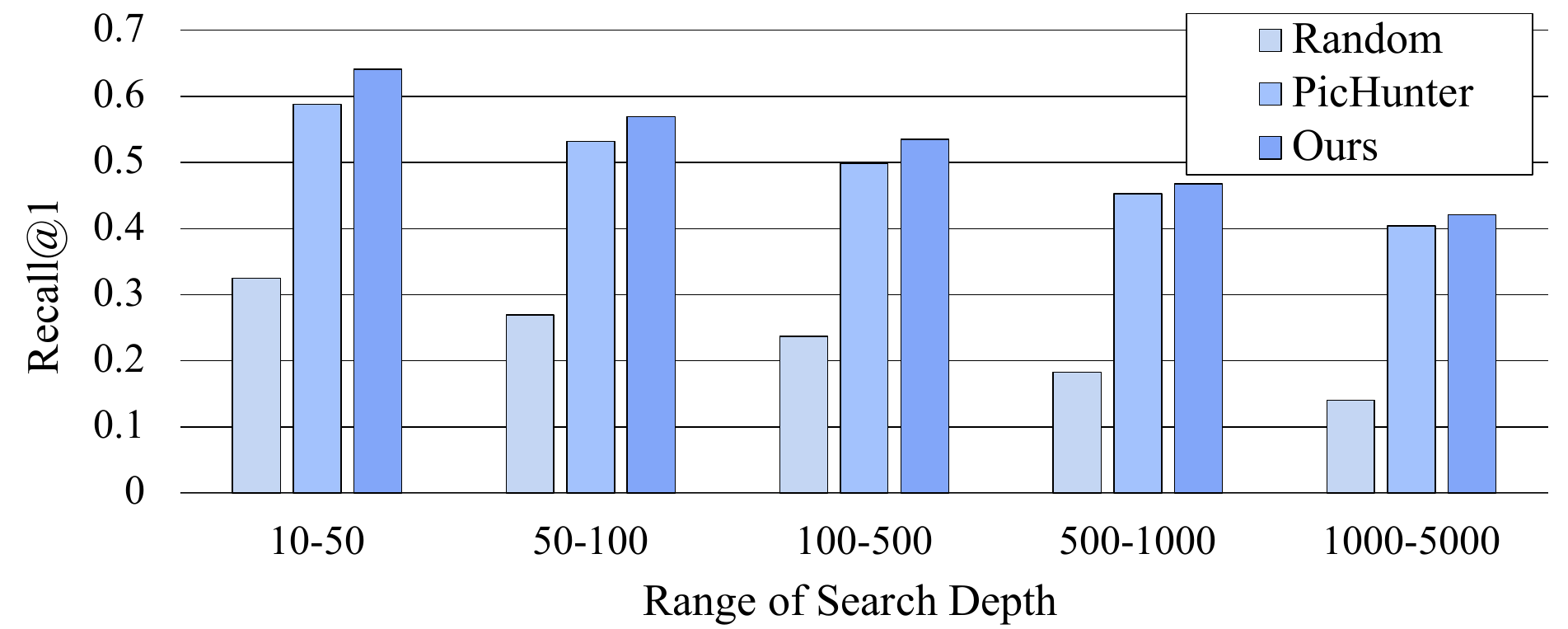}}
    \end{minipage}
    \caption{Recall@1 performance comparison.}
    \label{fig:exp_rangewise}
\end{figure}
% \vspace{-5px}

%% file: tabs/exp_prune.tex
% \begin{table}[h!]
% \caption{Recall@1 performance comparison at different steps with search space pruning.}
% \label{tab:robustrf_exp_prune}
% \centering
% \begin{tabular}{lccccccc}
% \hline
%  & Step-1 & Step-2 & Step-3 & Step-4 & Step-5 & Step-6 & Step-7 \\ \hline
% PicHunter & 0.0187    & 0.0507          & 0.1091          & 0.1874          & 0.2861 & 0.3942          & 0.4949          \\ 
%  - w/ Prune & 0.0202 & 0.0701 & 0.1646 & 0.2874 & 0.4147 & 0.5310 & 0.6234 \\ \cline{2-8} 
% Ours & 0.0238 & 0.0688 & 0.1409 & 0.2363 & 0.3438 & 0.4496 & 0.5467 \\ 
% - w/ Prune & \textbf{0.0240} & \textbf{0.0811} & \textbf{0.1835} & \textbf{0.3057} & \textbf{0.437} & \textbf{0.5491} & \textbf{0.6384} \\ 
% \hline \bottomrule
% \end{tabular}
% \end{table}

\vspace{-5px}

\begin{table}[h!]
\caption{Recall@1 performance comparison at different steps with search space pruning.}
\label{tab:exp_prune}
\centering
\begin{tabular}{lcccc}
\hline
 & Step-1 & Step-3 & Step-5 & Step-7 \\ \hline
PicHunter & 0.0187 & 0.1091 & 0.2861 & 0.4949 \\ 
- w/ Prune & 0.0202 & 0.1646 & 0.4147 & 0.6234 \\ \cline{2-5}
Ours & 0.0238 & 0.1409 & 0.3438 & 0.5467 \\ 
- w/ Prune & \textbf{0.0240} & \textbf{0.1835} & \textbf{0.437} & \textbf{0.6384} \\ 
\hline \bottomrule
\end{tabular}
\end{table}

\vspace{-10px}

%% file: tabs/exp_tkis_query.tex
\vspace{-5px}

\begin{table}[h]
    % \caption{Number of successfully identified t-KIS queries across various rounds (r1, r2, and r3). The query content is gradually completed from Round 1 to Round 3.}
    \caption{The performances on t-KIS queries. The results show the number of search targets successfully found when more details are included to these queries over three rounds.}  % (see Table~\ref{tab:tkis_text} in the supplementary paper)
    \label{tab:exp_tkis}
    \centering
    \begin{tabular}{lccc ccc}
        \toprule
        & \multicolumn{3}{c}{Recall@1} & \multicolumn{3}{c}{Recall@10} \\
        \cmidrule(lr){2-7}
        & r1 & r2 & r3 & r1 & r2 & r3 \\
        \cmidrule(lr){2-4} \cmidrule(lr){5-7}
        Random    & $6/17$  & $8/17$  & $8/17$  & $9/17$  & $12/17$  & $12/17$  \\
        PicHunter & $9/17$  & $11/17$ & $14/17$ & $11/17$ & $15/17$  & $16/17$  \\
        Ours      & $\textbf{10}/17$ & $\textbf{14}/17$ & $\textbf{16}/17$ & $\textbf{14}/17$ & $\textbf{17}/17$  & $\textbf{17}/17$  \\
        \hline \bottomrule
    \end{tabular}
\end{table}
        
\vspace{-15px}

%% file: conclusion.tex
\section{Conclusion}

Despite PicHunter as a classic model proposed more than 20 years ago, it can still effectively boost 40.05\% of search targets ranked at a depth beyond 1,000 to the top-1 rank within seven rounds of iteration on a million-scale video dataset. Nevertheless, such an impressive performance is grounded on the assumption that the user can provide relative judgment aligned with the machine's perception of the similarity measurement. By pairwise relative judgment and the predictive user model as proposed in this paper, we have relaxed PicHunter's assumption of a perfect user, allowing for imperfect user feedback by modeling user perception as a combination of multiple sub-perceptions. 
Our approach improves the robustness of PicHunter by boosting more search targets to top-1 across different search depths. At a depth beyond 1,000, our approach can boost 6.4\% of search targets to top-1. On VBS textual KIS search, our approach is also able to rank all the search targets to top-10, with the less number of iterations compared to PicHunter.
% Our approach improves robustness against inconsistent feedback and enhances search performance by progressively refining the search space based on the estimated alignment between user perception and feature representations.

% In PicHunter, when the assumed perfect user is not applicable, its user model is unable to filter inconsistent user feedback. Our method relaxes PicHunter's assumption, acknowledging that users are imperfect, but hypothesizing that human perception is likely to align with one or several of a set of feature representations. The simulated user makes judgments based on the majority decision across multiple sub-perceptions, with inconsistency arising when partial sub-perceptions disagree with the majority. We improve robustness against inconsistent feedback by designing a predictive user model that estimates the composition of user perception. The model-aligned user feedback is then used to progressively refine the search space. Experimental results suggest that PicHunter performs well when the majority of the sub-perceptions align with user perception, even though its user model cannot filter out inconsistent feedback. The proposed predictive user model enhances the system's robustness by identifying the sub-perceptions aligned with user perception, thereby improving search performance.

%% file: reference.bbl

%% file: appendix.tex
\clearpage
\appendix

\section{Ablation Study}

Tables~\ref{tab:exp_ablation} and \ref{tab:exp_ablation_prune} provide results from ablation studies that analyze the contribution of different components to the overall model performance. Table~\ref{tab:exp_ablation} represents the ablation results of the model in the PicHunter's setting, while Table~\ref{tab:exp_ablation_prune} includes results that are enhanced with search space pruning. In both tables, the performance of the proposed model (``Ours'') is compared with three ablated versions, each removing a specific component: ``SoftUpd'', ``StateRep'', ``DistEmb'' which denote the soft updating, the approximated state representation $v_{s^t}$, and the distance embedding $E^d$, respectively. Across all steps, the complete model consistently outperforms the ablated versions, demonstrating the effectiveness of these components. In Table~\ref{tab:exp_ablation}, after 7 steps, the full model achieves a Recall@1 of 0.5467, which is notably higher than the ablated models, particularly ``SoftUpd'' which achieves only 0.3321. Similarly, in Table~\ref{tab:exp_ablation_prune}, which includes search space pruning, the model with all components (``Ours w/ Prune'') achieves the highest Recall@1 values across all steps, reaching 0.6384 at Step-7, while ``SoftUpd'' and ``StateRep'' versions fall significantly behind. Notably, the performance gaps between the full model and the ablated versions are present in both settings, emphasizing the importance of all components in the model’s architecture. The enhanced version with pruning consistently yields superior performance compared to the original, further demonstrating the impact of search space pruning on improving the model's accuracy. These results suggest that each component plays a crucial role in maximizing model effectiveness, and their combined presence contributes to the overall performance gains in both the original setting and the one enhanced by search space pruning.

\input{tabs/exp_ablation}
\input{tabs/exp_ablation_prune}

\section{Effect of Distance Embedding}

As shown in Table~\ref{tab:exp_ablation}, the performance of ``- DistEmb'' is slightly better than ``Ours'' at the first step, with a margin of $0.0011$. In subsequent rounds, ``Ours'' gradually outperforms ``- DistEmb'', with the performance gap increasing from $0.0017$ to $0.0334$. These results support the discussion in Section~\ref{subsubsec:method_user_perception_prediction}: the distance embedding enables the model to capture subtle changes in $\|v_{\text{diff}}\|$, particularly in the later rounds. As described in Section~\ref{subsubsec:method_user_perception_prediction}, the range of possible $\|v_{\text{diff}}\|$ values, i.e., $(0, 1]$, is divided into 100 intervals, each associated with a learnable embedding. In the first round, $\|v_{\text{diff}}\|$ tends to be relatively large, but values in the higher end of the range are less frequent, which may result in under-trained embeddings. This likely contributes to the slightly lower performance observed in the first round.

\section{Textual KIS Query and User Feedback}

Table~\ref{tab:tkis_text} presents the query text for the VBS t-KIS from 2022 to 2024. As discussed in Section~\ref{sec:exp_tkis}, the full query is released in three rounds. We use colored solid, dashed, and dotted underlines to denote the query segments newly introduced in the first, second, and third rounds, respectively. Additionally, Figure~\ref{fig:feedback_example} provides an example of user feedback, specifically for query Q1 in Table~\ref{tab:tkis_text}. The user simulator applies majority voting based on the similarity measure of three sub-perceptions between the video pair and the search target. The predictive user model then filters out inconsistent sub-perceptions and refines the search target to achieve the top-1 position.

\input{figs/user_feedback_example}
\input{tabs/tkis_query}

%% file: tabs/exp_ablation.tex
% \begin{table*}[htp]
% \caption{Ablation study on the model components.}
% \label{tab:exp_ablation}
% \centering
% \begin{tabular}{lccccccc}
% \toprule
%             & Step-1    & Step-2          & Step-3          & Step-4          & Step-5          & Step-6          & Step-7          \\ \hline
% Ours        & 0.0238    & \textbf{0.0688} & \textbf{0.1409} & \textbf{0.2363} & \textbf{0.3438} & \textbf{0.4496} & \textbf{0.5467} \\
% - SoftUpd & 0.0196 & 0.0549 & 0.0993 & 0.1497 & 0.2113 & 0.2713 & 0.3321 \\
% - StateRep   & 0.0223    & 0.067           & 0.139           & 0.2304          & 0.3337          & 0.4349          & 0.5341          \\
% - DistEmb & \textbf{0.0249} & \textbf{0.0688} & 0.1392          & 0.2271          & 0.3271          & 0.424 & 0.5133 \\ \hline \bottomrule
% \end{tabular}
% \end{table*}

\begin{table}[h]
\caption{Ablation study on the model components.}
\label{tab:exp_ablation}
\centering
\begin{tabular}{lcccc}
\toprule
            & Step-1    & Step-3          & Step-5          & Step-7          \\ \hline
Ours        & 0.0238    & \textbf{0.1409} & \textbf{0.3438} & \textbf{0.5467} \\
- SoftUpd  & 0.0196    & 0.0993         & 0.2113         & 0.3321 \\
- StateRep   & 0.0223    & 0.139          & 0.3337          & 0.5341          \\
- DistEmb & \textbf{0.0249} & 0.1392          & 0.3271          & 0.5133 \\ \hline \bottomrule
\end{tabular}
\end{table}

%% file: tabs/exp_ablation_prune.tex
% \begin{table*}[h]
% \caption{Ablation study on the model components with search space pruning.}
% \label{tab:robustrf_exp_ablation_prune}
% \centering
% \begin{tabular}{lccccccc}
% \toprule
%             & Step-1    & Step-2          & Step-3          & Step-4          & Step-5          & Step-6          & Step-7          \\ \hline
% Ours w/ Prune & 0.0240 & \textbf{0.0811} & \textbf{0.1835} & \textbf{0.3057} & \textbf{0.4370} & \textbf{0.5491} & \textbf{0.6384} \\
% - SoftUpd & 0.0169 & 0.0391 & 0.0677 & 0.1029 & 0.1429 & 0.1847 & 0.2255 \\
% - StateRep & \textbf{0.0253} & 0.0794 & 0.1762 & 0.3030 & 0.4315 & 0.5401 & 0.6311 \\
% - DistEmb & 0.0244 & 0.0805 & 0.1770 & 0.2997 & 0.4287 & 0.5387 & 0.6291 \\
% \hline \bottomrule
% \end{tabular}
% \end{table*}

\begin{table}[h]
\caption{Ablation study on the model components with search space pruning.}
\label{tab:exp_ablation_prune}
\centering
\begin{tabular}{lcccc}
\toprule
            & Step-1    & Step-3          & Step-5          & Step-7          \\ \hline
Ours w/ Prune & 0.0240 & \textbf{0.1835} & \textbf{0.4370} & \textbf{0.6384} \\
- SoftUpd   & 0.0169 & 0.0677         & 0.1429         & 0.2255 \\
- StateRep  & \textbf{0.0253} & 0.1762         & 0.4315         & 0.6311 \\
- DistEmb   & 0.0244 & 0.1770         & 0.4287         & 0.6291 \\
\hline \bottomrule
\end{tabular}
\end{table}

%% file: figs/user_feedback_example.tex
\vspace{-10px}
\begin{figure}[hbp]
    \center
    \begin{minipage}{1. \columnwidth}
        \centerline{\includegraphics[width=1. \columnwidth]{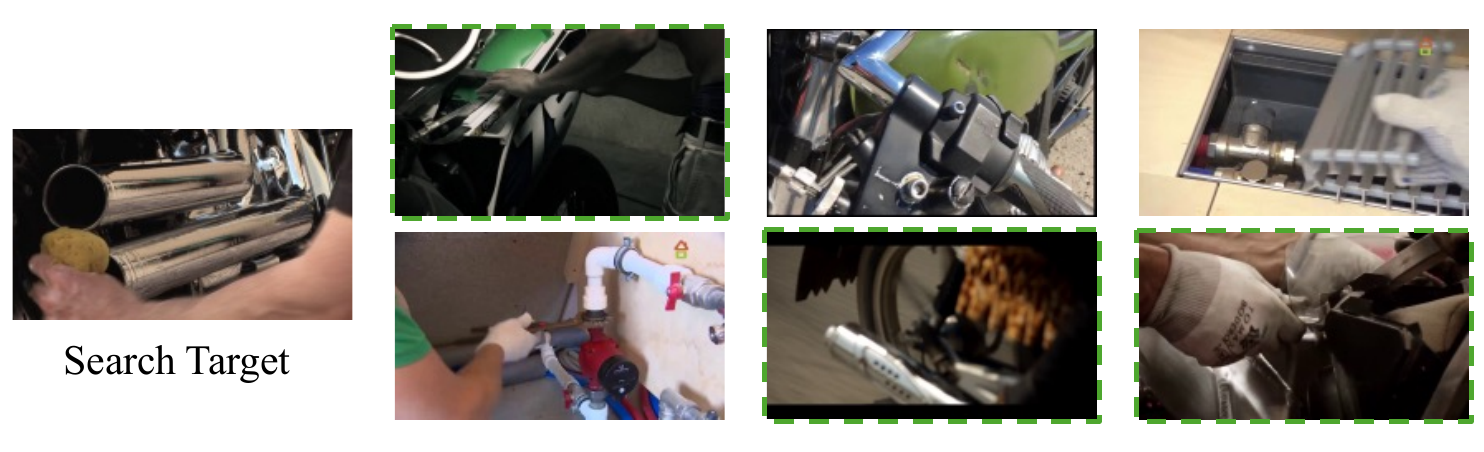}}
    \end{minipage}
    \caption{An example of user feedback. Given the search target on the left, the user simulator provides feedback on the pairs, with the green dashed boundary indicating selection. The user model then filters out irrelevant sub-perceptions and promotes the search target to the top-1 position. The initial textual query is: ``Close-up of motorbike exhaust pipes being cleaned with a wet sponge. Two chromed pipes are visible, open on the left.''}
    \label{fig:feedback_example}
\end{figure}

%% file: tabs/tkis_query.tex
\begin{table*}[ht]

    \caption{The query text of the t-KIS from VBS 2022 to 2024. The query text is divided into three parts which are visually distinguished using colored solid, dashed, and dotted underlines. In each round, the query is presented to the participants and is appended to the previous text.}
    \label{tab:tkis_text}
    
    \centering
    \begin{tabular}{>{\centering\arraybackslash}m{1.2cm} >{\raggedright\arraybackslash}m{14cm}}
    
    \hline
    \textbf{Query ID} & \textbf{Textaul KIS Query} \\
    \hline
    Q1 & 
    {\color{red}\uline{Close-up of motorbike exhaust pipes being cleaned with a wet sponge.}}
    {\color{cyan}\dashuline{Two chromed pipes are visible, open on the left.}}
    {\color{blue}\dotuline{The forearm of a man wearing black T-Shirt is visible a few times.}} \\
    \cline{2-2}
    Q2 & 
    {\color{red}\uline{Almost static shot of a brown-white caravan and a horse on a meadow.}}
    {\color{cyan}\dashuline{The caravan is in the center, the horse in the back to its right, and there is a large tree on the right.}}
    {\color{blue}\dotuline{The camera is slightly shaky, and there is a forested hill in the background.}} \\
    \cline{2-2}
    Q3 & 
    {\color{red}\uline{Shot of an opened magazine, showing a drawing of a bearded man on the right side, then a shot of a person standing in a street and holding different pages of an open magazine in front of the camera.}}
    {\color{cyan}\dashuline{The person in the street wears a blue T-shirt and light grey jacket, and is wearing a mask and sunglasses.}}
    {\color{blue}\dotuline{There are white frames with black text messages flashing up in between. The drawing in the first shot is on black background, the man has a white beard, the title of the left page is "vote for Pedro".}} \\
    \cline{2-2}
    Q4 & 
    {\color{red}\uline{A shot of a man in a water slide, followed by two shots of two men trying to light a fire on a beach.}}
    {\color{cyan}\dashuline{The man slides down head first, and wears black bathing trunks.}}
    {\color{blue}\dotuline{There is a circle of stones around the fire, and we do not see the heads of the two men.}} \\
    \cline{2-2}
    Q5 &
    {\color{red}\uline{View from an upper deck of a ship down to a lower deck and water, slowly changing the view to the front of the ship, where a man with a camera walks into view.}}
    {\color{cyan}\dashuline{The lower deck is on the left, with green floor and two red/orange chairs, and water is on the right.}}
    {\color{blue}\dotuline{The man wears black trousers and a grey jacket.}} \\
    \cline{2-2}
    Q6 &
    {\color{red}\uline{Kids in kayaks on a river, throwing paddles through three coloured hoops placed over the water.}}
    {\color{cyan}\dashuline{The sequence starts with two kids next to each other in red kayaks with red paddles.}}
    {\color{blue}\dotuline{The river is partly lined with trees and there are spectators on the accessible spots. We see shots from both sides of the river.}} \\
    \cline{2-2}
    Q7 &
    {\color{red}\uline{Close-up shots of making coffee: grinding beans manually, putting the powder into a French press, and pouring coffee into a red mug.}}
    {\color{cyan}\dashuline{The French press is also red, and we see it operated from the top.}}
    {\color{blue}\dotuline{Pouring beans into the grinder is also shown from the top. The red mug has the text "The Pursuit" in a black circle on it.}} \\
    \cline{2-2}
    Q8 &
    {\color{red}\uline{Close-up of a man lighting a match on a stone, and then cooking at a fireplace in the woods.}}
    {\color{cyan}\dashuline{He uses a black pot with water and pours powder into it while stirring.}}
    {\color{blue}\dotuline{The fireplace is surrounded by stones, and we see a blue tent in the background.}} \\
    \cline{2-2}
    Q9 &
    {\color{red}\uline{Hands of a kid applying glue to an egg carton and then a view of a sculpture made of those cartons.}}
    {\color{cyan}\dashuline{In the second shot, the camera pans up along green and turquoise egg cartons.}}
    {\color{blue}\dotuline{On the first shot, we see a jar with white glue, the bottom of an egg carton and the kid holding a brush.}} \\
    \cline{2-2}
    Q10 &
    {\color{red}\uline{A two-masted sailing ship leaves a harbour, moving right behind a stone wall and a white light beacon.}}
    {\color{cyan}\dashuline{There are mountains in the background, and the sun is behind clouds.}}
    {\color{blue}\dotuline{There are a number of smaller boats in front, one of them flying a Turkish flag.}} \\
    \cline{2-2}
    Q11 &
    {\color{red}\uline{A sequence of three shots: two people and a wall with posters, a balcony with laundry hanging on a rope and a train passing behind two standing tombstones.}}
    {\color{cyan}\dashuline{In the first shot, a person is standing around a corner of the wall on the right, the other person walks away to the left.}}
    {\color{blue}\dotuline{In the last shot, there is a fence between the graveyard and the railway line, and the roof of a building is visible behind the railway line.}} \\
    \cline{2-2}
    Q12 &
    {\color{red}\uline{View down the surface of a boulder, with a forest in the background.}}
    {\color{cyan}\dashuline{A bearded man in a cyan shirt climbing up the boulder.}}
    {\color{blue}\dotuline{It is sunny, and we see the man's shadow on a smaller boulder on the right. The man wears gloves, but does not use any ropes.}} \\
    \cline{2-2}
    Q13 &
    {\color{red}\uline{View down from the helmet camera of a mountain biker, as he spins around on a path along a narrow ridge.}}
    {\color{cyan}\dashuline{He spins by jumping on the back wheel.}}
    {\color{blue}\dotuline{The ridge is flanked by sea. We hear the biker narrating the scene.}} \\
    \cline{2-2}
    Q14 &
    {\color{red}\uline{A girl and a man run up a small hill.}}
    {\color{cyan}\dashuline{There is a flagpole with a Canadian flag on top.}}
    {\color{blue}\dotuline{There are white/green benches, a lamppost and stairs up the hill. The flagpole is on top of a stone structure with a cannon.}} \\
    \cline{2-2}
    Q15 &
    {\color{red}\uline{A herd of donkeys/mules are walking down a walkway with steps, followed by a herdsman.}}
    {\color{cyan}\dashuline{The animals are saddled, with blankets in different color.}}
    {\color{blue}\dotuline{The herdsman wears a bonnet and carries a stick.}} \\
    \cline{2-2}
    Q16 &
    {\color{red}\uline{A man, leaning forward, looks back over his shoulder, where a smiling bride is walking towards him.}}
    {\color{cyan}\dashuline{The bride puts her arm around his neck.}}
    {\color{blue}\dotuline{The scene is outdoors, the man wears a striped shirt, black jacket and glasses with black/gray frame.}} \\
    \cline{2-2}
    Q17 &
    {\color{red}\uline{We see a girl in a dark dress pushing the door of a convenience store, after it closes, she runs away.}}
    {\color{cyan}\dashuline{There are two bikes and four trash cans in front of the shop windows.}}
    {\color{blue}\dotuline{The store's brand colors are green, white and blue.}} \\
    \hline
\end{tabular}
\end{table*}